# Merkle Trees in Blockchain: A Study of Collision Probability and Security Implications

Oleksandr Kuznetsov [1,2*], Alex Rusnak [1], Anton Yezhov [1], Kateryna Kuznetsova [1], Dzianis Kanonik [1], Oleksandr Domin [1]

[1] Proxima Labs, 1501 Larkin Street, suite 300, San Francisco, USA
[2] Department of Political Sciences, Communication and International Relations, University of Macerata, Via Crescimbeni, 30/32, 62100 Macerata, Italy

*Corresponding author. E-mail(s): kuznetsov@karazin.ua

**Abstract:** In the rapidly evolving landscape of blockchain technology, ensuring the integrity and security of data is paramount. This study delves into the security aspects of Merkle Trees, a fundamental component in blockchain architectures, such as Ethereum. We critically examine the susceptibility of Merkle Trees to hash collisions, a potential vulnerability that poses significant risks to data security within blockchain systems. Despite their widespread application, the collision resistance of Merkle Trees and their robustness against preimage attacks have not been thoroughly investigated, leading to a notable gap in the comprehensive understanding of blockchain security mechanisms. Our research endeavors to bridge this gap through a meticulous blend of theoretical analysis and empirical validation. We scrutinize the probability of root collisions in Merkle Trees, considering various factors such as hash length and path length within the tree. Our findings reveal a direct correlation between the increase in path length and the heightened probability of root collisions, thereby underscoring potential security vulnerabilities. Conversely, we observe that an increase in hash length significantly reduces the likelihood of collisions, highlighting its critical role in fortifying security. The insights garnered from our research offer valuable guidance for blockchain developers and researchers, aiming to bolster the security and operational efficacy of blockchain-based systems.

**Keywords:** Blockchain Security, Merkle Trees, Hash Collisions, Data Integrity, Blockchain Data Verification, Collision Resistance.

## 1. Introduction

In the era of digitalization and the rise of blockchain technologies, issues of data security and verification have become increasingly critical [1,2]. One of the key tools in this domain is Merkle Trees, widely utilized in blockchain systems such as Ethereum [3,4]. Merkle Trees facilitate the efficient proof of data inclusion within datasets, employing mechanisms of cryptographic hashing [5,6]. The core concept involves creating a data structure where each leaf represents a hash of preceding leaves, and the root of the tree (Merkle Root) serves as an aggregated representation of the entire dataset [7–9].

Merkle Proof is a mechanism that allows for the verification of the presence of a specific data element in a Merkle Tree without requiring access to the entire tree. This is achieved by providing a path from the leaf to the root, enabling any system participant to independently compute and verify the hash of the data of interest [5,10]. If the computed hash matches the Merkle Root, it confirms the data's presence in the tree.

These mechanisms not only offer simplicity and efficiency in data verification but also play a pivotal role in ensuring the security of blockchain systems [11,12]. The hashing functions used in blockchains must be resistant to collisions and preimage attacks, which is critically important for preventing data tampering and potential security threats [3,13].

However, despite the widespread use of Merkle Trees in blockchain technologies, questions related to the security of these structures, particularly their resistance to collisions and preimage attacks, remain insufficiently explored. This represents a significant gap in understanding and assessing the security of key components of blockchain systems.

Our article aims to fill this gap. We focus on assessing the probability of root collisions in Merkle Trees for various input data. This aspect holds significant practical importance as it allows for the evaluation of the security and reliability of Merkle Tree and Merkle Proof as fundamental elements of modern blockchain systems. The results of our research provide valuable insights for developers and researchers in the blockchain field, contributing to the enhancement of the security and efficiency of these technologies.

Our research article is structured as follows: We begin with a "State of the Art" section, reviewing existing literature and identifying research gaps. The "Methods" section is divided into two parts, discussing "Cryptographic Hash Functions" and "Merkle Trees," and highlighting our unique contribution to the field. In the "Results" section, we present our theoretical analysis of collision probabilities in Merkle Trees, including the "Birthday Paradox Strategy," and detail our experimental studies with methodology and results. The article then moves to the "Discussion of Research Findings," where we interpret the implications of our findings. Finally, the "Conclusion" summarizes the key insights and contributions of our research, emphasizing its impact on blockchain security and future research potential.

## 2. State of the Art

The rapid advancement of blockchain technology has brought to the forefront the critical importance of secure and efficient data verification mechanisms. Among these, Merkle trees play a pivotal role, offering a robust structure for data integrity and authentication. However, the evolving landscape of blockchain technology and its associated security challenges necessitate continuous research and development. This section reviews recent scholarly contributions in the field, highlighting their advancements and identifying gaps that our research aims to fill.

Backes and his team [14] provide machine-checked proofs of collision-resistance in Merkle-Damgård constructions. While their work significantly contributes to the theoretical security analysis of hash functions, it does not delve into the practical aspects of Merkle tree implementations in blockchain technologies.

Andreeva and colleagues [15] present new generic second-preimage attacks on various Merkle-Damgård-based hash functions. Their findings offer a deeper understanding of the vulnerabilities in hash functions but do not specifically address the impact of these vulnerabilities on the overall security of Merkle trees in blockchain systems.

Bao and colleagues [16] explore the security of hash combiners, revealing that the security of most combiners is not as high as commonly believed. Their research is pivotal in understanding the limitations of hash function combinations but stops short of examining the specific role of these combiners in the structure and security of Merkle trees in blockchain applications.

Diván and Sánchez-Reynoso [10] focus on metadata-based measurements transmission verified by a Merkle Tree, emphasizing data stream processing strategies. Their contribution lies in optimizing data transmission and integrity verification, yet the study primarily centers on data transmission efficiency rather than the broader security implications in blockchain networks.

In their work on time-space lower bounds for finding collisions in Merkle-Damgård hash functions, Akshima, Guo, and Liu [17] explore the computational complexity of collision attacks in the auxiliary-input random oracle model. While providing valuable insights into the security of Merkle-Damgård hash functions, their research primarily focuses on the theoretical aspects of hash function vulnerabilities, leaving practical implications in real-world blockchain systems less explored.

Al-Odat, Khan, and Al-Qtiemat [18] propose an enhanced secure hash design to mitigate collision and length extension attacks in SHA-1 and SHA-2. Their work contributes to improving the resilience of hash functions against specific types of attacks. However, the study does not directly address the broader implications of these vulnerabilities in the context of Merkle trees within blockchain environments.

Our research fills a critical gap by specifically focusing on the probability of root collisions in Merkle trees and their implications for blockchain security. We extend the existing body of knowledge by not only addressing theoretical vulnerabilities but also by providing practical insights and solutions for enhancing the security and efficiency of blockchain systems.

## 3. Methods

This section succinctly outlines the fundamental techniques underpinning blockchain technology: Cryptographic Hash Functions and Merkle Trees. It provides a concise overview of these key components, essential for ensuring data integrity and security in blockchain systems, highlighting their role in efficiently managing complex data structures within blockchain architectures.

### 3.1. Cryptographic Hash Functions

Cryptographic hash functions are integral to the security infrastructure of digital systems, particularly in blockchain technology. They serve as the backbone for ensuring data integrity and authentication, playing a pivotal role in various cryptographic operations.

A cryptographic hash function, denoted as $H$, is a mathematical algorithm that maps an input $x$ of arbitrary length to a fixed-size string of bytes, typically referred to as the hash value $h$, such that $H(x) = h$ [19]. The transformation aims to be quick, deterministic, and irreversible, ensuring data integrity and security.

The key cryptographic properties of hash functions include [19]:
- Pre-image Resistance: For any given hash output $h$, it should be computationally infeasible to find any input $x$ such that $H(x) = h$.
- Second Pre-image Resistance: For any given input $x$, it should be computationally impractical to find a different input $x'$ such that $H(x) = H(x')$.
- Collision Resistance: It should be computationally unfeasible to find any two distinct inputs $x$ and $x'$ that hash to the same output, i.e., $H(x) = H(x')$.
- Uniform Distribution and Avalanche Effect: The hash function should uniformly distribute the hash values over its output space and ensure that a small change in the input significantly alters the output. Mathematically, if $x$ and $x'$ differ by only a few bits, then $H(x)$ and $H(x')$ should be significantly different.
- Determinism: The function should be deterministic, implying that the same input will always produce the same hash output.
- Efficiency: The function should compute the hash value efficiently, making it practical for real-time applications. This implies a low computational complexity, typically polynomial in the size of the input.

These properties ensure that cryptographic hash functions can securely convert data of arbitrary size into a fixed-size hash, which can then be used in various applications, including digital signatures, message integrity checks, and the construction of Merkle trees in blockchain systems.

### 3.2. Merkle Trees

Merkle Trees, named after their inventor Ralph Merkle [20,21], are a fundamental concept in computer science and cryptography, particularly in the domain of blockchain technology. These tree structures play a crucial role in efficiently summarizing and verifying large datasets, such as transactions in a blockchain [11,12].

A Merkle Tree is a binary tree in which every leaf node contains the hash of a data block, and every non-leaf node contains the cryptographic hash of the concatenation of its child nodes [20,21]. For data blocks $d_1, d_2, ..., d_n$, the corresponding hashes $h_1, h_2, ..., h_n$ are computed as $h_i = H(d_i)$. Each non-leaf node $h_i$ is then defined as $h_{ij} = H(h_i \| h_j)$, where $\|$ denotes concatenation. This structure allows for efficient and secure verification of the contents of large data sets.

The root of the tree, denoted as $r$, is computed recursively by combining the hashes of its child nodes, expressed as:
$$r = r(d_i) = H(H(...H(H(H(d_i) \| h_1) \| ...) \| h_{k-1}) \| h_k).$$

The Merkle Path is the sequence of hashes $h = \{h_1, h_2, ..., h_k\}$ from a leaf node $d_i$ to the root $r$, which is used to verify the presence of a specific data block. A Merkle Proof consists of a data block $d_i$, its corresponding Merkle Path $h = \{h_1, h_2, ..., h_k\}$, and the root hash $r$. Verifying the proof involves hashing the data block $d_i$, successively concatenating and hashing it with the hashes in the Merkle Path, and comparing the result with the root hash $r$.

Key properties of Merkle Trees include [20,21]:
- Efficient Verification: Merkle Trees enable efficient verification of data integrity. Given a root hash $r$, one can verify the presence of a specific data block $d_i$ in the tree without needing to check every node.
- Tamper-Evident: Any alteration in a data block $d_i$ alters the hash stored in the leaf node, which cascades up to the root $r$ of the tree, making any tampering evident.
- Space Efficiency: Merkle Trees summarize a large dataset $d_1, d_2, ..., d_n$ in a much smaller cryptographic footprint (the root hash $r$), saving space and facilitating easy transmission of proof.

In blockchain systems, Merkle Trees are used to efficiently summarize all transactions in a block. Each transaction is a leaf node $d_i$ in the tree. This structure allows users to verify whether a particular transaction is included in a block without downloading the entire blockchain. For instance, in Bitcoin, Merkle Trees are used to create a compact summary of all transactions in a block, enabling lightweight nodes to verify transactions independently (Fig. 1). This hierarchical structure facilitates efficient data processing and verification, aligning with the computing approach of handling complex information entities.

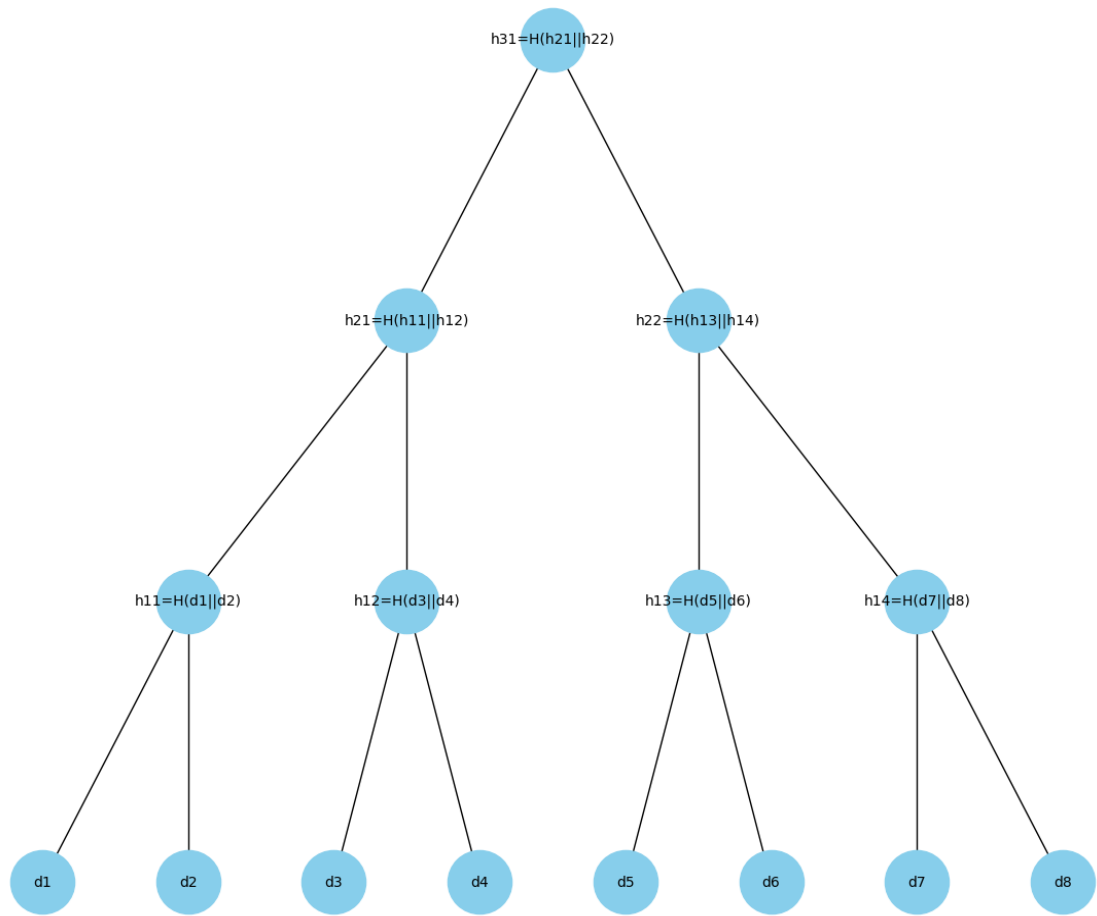

Fig.1. Merkle tree structure

In summary, Merkle Trees are a vital component in the architecture of blockchain systems, providing a secure and efficient means of data verification.

**Our contribution to the field of research**

Our research addresses a notable gap in the existing studies on the security of Merkle Trees, particularly in the context of blockchain technology. While previous research has extensively explored the cryptographic properties of hash functions and the theoretical aspects of Merkle Trees, there is a lack of comprehensive analysis on the probability of root collisions in Merkle Trees and their implications for the overall security and integrity of blockchain systems.

This gap is critical because root collisions in Merkle Trees could potentially undermine the security mechanisms of blockchain networks, leading to vulnerabilities in data integrity and authentication processes. Our study aims to fill this gap by providing a detailed analysis of the likelihood of such collisions and offering insights into enhancing the robustness of Merkle Trees in blockchain applications. This includes examining the impact of various parameters like hash length and tree depth on the security of Merkle Trees, thereby contributing to the development of more secure and efficient blockchain architectures.

### 4. Results

In this section, we present the key findings of our research, encompassing both theoretical insights and experimental data that corroborate these theoretical premises. We begin with a detailed analysis of the theoretical aspects of estimating collision probabilities in Merkle Trees, based on the assumption of using an ideal cryptographic hash function. Subsequently, we will discuss the outcomes of experiments conducted to validate and confirm these theoretical conclusions.

## 4.1 Theoretical Analysis of Collision Probabilities in Merkle Trees

Let us assume that for hashing data $d_i$, an ideal cryptographic hash function $H$ is employed, generating a hash code $H(d_i)$ of length $m$ bits. In the context of our study, this ideal cryptographic hash function is a fundamental element that ensures the security and reliability of Merkle Trees. Such a function is characterized by its resistance to collisions, implying an extremely low probability that two different inputs will yield the same hash code. Ideally, every possible hash code of length $m$ bits is generated with an equal probability of $2^{-m}$, rendering it practically impossible to deliberately select inputs that lead to a specific hash code. This property is critically important for ensuring the integrity and security of data in blockchain systems, where Merkle Trees are utilized for efficient and reliable data verification.

Using an ideal cryptographic hash function, the probability that the outputs for two different inputs $d_i \neq d_j$ will coincide, $H(d_i) = H(d_j)$, is given by:
$$P(H(d_i) = H(d_j)) = 2^{-m}.$$

Now, let us consider a Merkle Tree with root $r$.

For arbitrary input data $d_i$ and associated Merkle path $h = \{h_1, h_2, ..., h_k\}$, the root $r$ is calculated as follows:
$$r = r(d_i) = H(H(...H(H(H(d_i) \| h_1) \| ...) \| h_{k-1}) \| h_k).$$

We estimate the probability $P(r = r')$ that for two different inputs $d_j \neq d_i$, the corresponding roots coincide, $r = r'$, where
$$r' = r(d_j) = H(H(...H(H(H(d_j) \| h_1) \| ...) \| h_{k-1}) \| h_k).$$

Considering the case where the Merkle path length $k = 1$, we have:
$$r = r(d_i) = H(H(d_i) \| h_1), \quad r' = r(d_j) = H(H(d_j) \| h_1).$$

The equality $r = r'$ is achieved in two scenarios:
- When $H(d_i) = H(d_j)$, indicating a direct hash match of the input data.
- Or, when $H(d_i) \neq H(d_j)$ and $H(H(d_i) \| h_1) = H(H(d_j) \| h_1)$, indicating a hash match after the first level of hashing in the Merkle Tree.

Given that the probability of hash matches for different inputs is
$$P(H(d_i) = H(d_j))|_{d_j \neq d_i} = P(H(H(d_i) \| h_1) = H(H(d_j) \| h_1))|_{H(d_i) \neq H(d_j)} = 2^{-m},$$
we can express the probability $P(r = r')$ as:
$$P(r = r') = 2^{-m} + (1 - 2^{-m})2^{-m}.$$

Moving to the case where $k = 2$, we have:
$$r = r(d_i) = H(H(H(d_i) \| h_1) h_2)$$
and
$$r' = r(d_j) = H(H(H(d_j) \| h_1) \| h_2).$$

Here, the equality $r = r'$ can also be achieved in several scenarios:
- When $H(d_i) = H(d_j)$,
- When $H(H(d_i) \| h_1) = H(H(d_j) \| h_1)$ given $H(d_i) \neq H(d_j)$,
- Or when $H(H(H(d_i) \| h_1) \| h_2) = H(H(H(d_j) \| h_1) \| h_2)$ given $H(d_i) \neq H(d_j)$ and $H(H(d_i) \| h_1) \neq H(H(d_j) \| h_1))$.

Thus, the probability $P(r = r')$ for $k = 2$ is:
$$P(r = r') = 2^{-m} + (1 - 2^{-m})2^{-m} + (1 - 2^{-m})^2 2^{-m}.$$

Generalizing for an arbitrary path length $k$, we derive the following formula to estimate the probability of root collisions in Merkle Trees:

$$P(r=r') = 2^{-m} + (1-2^{-m})2^{-m} + (1-2^{-m})^2 2^{-m} + \ldots + (1-2^{-m})^k 2^{-m} = 2^{-m} + 2^{-m}\sum_{i=1}^{k}(1-2^{-m})^i.$$

This formula reflects the probability of root collisions in Merkle Trees as a function of the path length $k$ and hash length $m$, demonstrating how the likelihood of collisions increases with the length of the path.

For a more compact representation and analysis of the formula for $P(r=r')$, let's consider the sum of the first $k$ terms of a geometric progression:

$$S = \sum_{i=1}^{k}(1-2^{-m})^i.$$

This sum is a classic mathematical problem, the solution of which allows us to efficiently compute the overall probability $P(r=r')$. Viewing $S$ as the sum of a geometric progression, we obtain:

$$\sum_{i=1}^{k}(1-2^{-m})^i = (1-2^{-m})\frac{1-(1-2^{-m})^k}{1-(1-2^{-m})} = (2^m-1)(1-(1-2^{-m})^k).$$

Substituting this expression into the formula for $P(r=r')$, we arrive at the following result:

$$P(r=r') = 2^{-m} + 2^{-m}(2^m-1)(1-(1-2^{-m})^k) = 2^{-m} + (1-2^{-m})(1-(1-2^{-m})^k). \quad (1)$$

This formula represents a key outcome of our theoretical analysis. It demonstrates that the probability of root collisions in Merkle Trees is not static but depends on the path length $k$ and the hash size $m$.

Figure 1 presents a three-dimensional graph showing the dependency of $P(r=r')$ on parameters $k$ and $m$. Figure 2 illustrates the same dependency in a logarithmic scale for the most practically relevant cases with $k=0,\ldots,64$ and $m=128,\ldots,256$. The dependencies shown in Figures 1 and 2 were calculated using formula (1), and for visualization, the professional computer algebra system Wolfram Mathematica was employed.

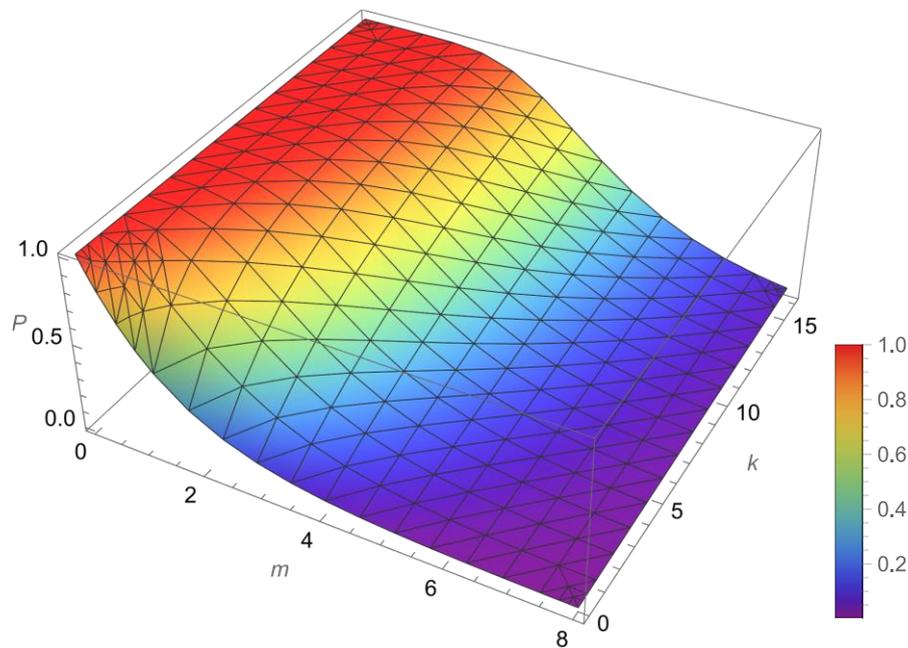

Figure 1 – Dependency of the probability $P(r=r')$ on parameters $k$ and $m$

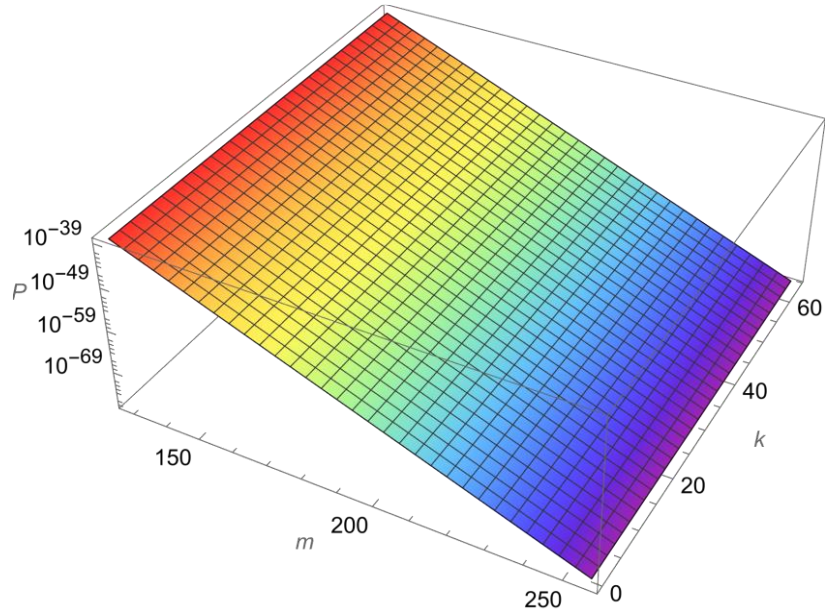

Figure 2 – Dependency of the probability $P(r=r')$ in logarithmic scale for parameters $k=0,...,64$ and $m=128,...,256$

It's important to note that as $k$ increases, the probability of root collisions in Merkle Trees also increases, potentially diminishing the security of Merkle Proofs. This significant finding has substantial implications for the security of blockchain systems utilizing Merkle Trees, and it's crucial to assess the magnitude of this threat.

To simplify the computation and analysis of expression (1), we utilize the properties of Taylor series. The Taylor series allows a function to be represented as an infinite sum of its derivatives, calculated at a specific point. For the exponential function $e^x$, the Taylor series around point 0 is:

$$e^x = 1 + x + \frac{x^2}{2!} + \frac{x^3}{3!} + \cdots.$$

In our research context, for large values of $m$, the value $2^{-m}$ becomes so small that we can use the approximation of the first terms of the Taylor series for $e^{-x}$, where $x = -2^{-m}$. Thus, the approximation can be expressed as:

$$e^{-2^{-m}} \approx 1 - 2^{-m}.$$

This approximation is based on the fact that for small values of $x$, higher powers of $x$ in the Taylor series become insignificant, and therefore the first terms of the series provide sufficient accuracy.

Applying this approximation to the probability of root collisions, we obtain:

$$P(r=r') = 2^{-m} + (1-2^{-m})(1-(1-2^{-m})^k) \approx 2^{-m} + e^{-2^{-m}}(1-e^{-k2^{-m}}) = 2^{-m} + e^{-2^{-m}} - e^{-(k+1)2^{-m}}. \quad (2)$$

The last formula allows us to estimate $P(r=r')$ in the case of extremely large values of $k$. For example, for $k = 2^m$, formula (2) simplifies to:

$$P(r=r') \approx 2^{-m} + e^{-2^{-m}}(1-e^{-2^m 2^{-m}}) = 2^{-m} + e^{-2^{-m}}\left(1-\frac{1}{e}\right).$$

For large $m$, we have $e^{-2^{-m}} \approx 1$ and $2^{-m} \approx 0$, leading to:

$$P(r=r') \approx 1 - \frac{1}{e} \approx 0,63.$$

Thus, increasing the path length in Merkle Trees inevitably leads to a higher probability of collisions, reaching 63% when $k = 2^m$. This is clearly visible in Figure 1, where, for instance, for $m = 4$ and large values of $k$, the probability $P(r = r')$ indeed exceeds 50% and, evidently, reaches 0.63 for $k = 2^m = 16$. This crucial insight serves as a warning and necessitates a reevaluation of the parameters used in security mechanisms in modern applications, especially in the context of blockchain systems and cryptocurrencies. The use of Merkle Trees may only be justified in cases of small values of $k$ relative to $2^m$. The situation could be significantly exacerbated in the event of a so-called "birthday attack."

**The Birthday Paradox Strategy**

Thus far, we have considered scenarios where, for a fixed Merkle path $h = \{h_1, h_2, ..., h_k\}$ and root $r$, it is possible to find different data $d_i$ that lead to a collision. This means the computed root $r'$ for data $d_j \neq d_i$ and path $h$ matches the root $r$. Practically, this would imply substituting true data $d_i$ with false data $d_j \neq d_i$, where the Merkle proof mechanism fails to detect the forgery and erroneously verifies it as authentic information. This is a highly undesirable scenario, particularly for cryptocurrencies or other blockchain applications, where it poses a real threat of digital asset loss. However, this threat can be further amplified by considering another equally intriguing scenario.

Let's now consider a hypothetical case where we can select not only the data $d_j \neq d_i$ but also the root $r$ (and possibly the paths $h = \{h_1, h_2, ..., h_k\}$, $h' = \{h'_1, h'_2, ..., h'_k\}$), i.e., we want to find any possible combination of $d_i, d_j, r, h, h'$ that results in the equality:

$$r = r(d_i) = H(H(...H(H(H(d_i) \| h_1) \| ...) \| h_{k-1}) \| h_k) =$$
$$= r' = r'(d_j) = H(H(...H(H(H(d_j) \| h'_1) \| ...) \| h'_{k-1}) \| h'_k).$$

First, let's consider the simplest case:
$$r = r(d_i) = H(d_i) =$$
$$= r' = r'(d_j) = H(d_j).$$

We can denote this as a Merkle Tree variant with $k = 0$ (essentially, we are comparing the hashing results of two different data $d_j \neq d_i$ without using a Merkle path).

The task of finding a pair $d_j \neq d_i$ that satisfies $H(d_i) = H(d_j)$ is described by the well-known "birthday paradox" [22]. This paradox states that the probability that two people in a group have the same birthday becomes higher than 50% if the group has 23 or more people [23]. This is counterintuitive, as the number of days in a year is much larger.

In the context of hash functions $H$, where the hash code $h_i = H(d_i)$ has $m$ bits, the number of possible unique hash codes is $2^m$. The probability that two random inputs do not lead to a collision is $1 - \frac{1}{2^m}$ for the first pair. For the second pair, this probability is $1 - \frac{2}{2^m}$, as there is already one selected hash code, and so on. After selecting $s$ hash codes, the probability that the next hash code does not lead to a collision is $1 - \frac{s-1}{2^m}$.

Therefore, the probability that there will be no collision after selecting $s$ hash codes is the product of all these probabilities:

$$P(H(d_i) \neq H(d_j)) = \left(1 - \frac{1}{2^m}\right) \times \left(1 - \frac{2}{2^m}\right) \times ... \times \left(1 - \frac{s-1}{2^m}\right) = \prod_{i=0}^{s-1}\left(1 - \frac{i}{2^m}\right). \quad (3)$$

Then, the probability of a collision is

$$P(H(d_i) = H(d_j)) = 1 - P(H(d_i) \neq H(d_j)) = 1 - \prod_{i=0}^{s-1}\left(1 - \frac{i}{2^m}\right).$$

The product of probabilities used in (3) can be approximated using the second remarkable limit. This limit states that

$$\lim_{n \to \infty}\left(1 + \frac{1}{n}\right)^n = e,$$

where $e$ is the base of the natural logarithm.

For large values of $m$ and relatively small $n$, the value $\frac{i}{2^m}$ is very small. Consequently, we can use the approximation $e^{\frac{-i}{2^m}} \approx 1 - \frac{i}{2^m}$:

$$P(H(d_i) \neq H(d_j)) = \prod_{i=0}^{s-1}\left(1 - \frac{i}{2^m}\right) \approx \prod_{i=0}^{s-1} e^{\frac{-i}{2^m}}.$$

The product of exponents equals the exponent of the sum, therefore:

$$P(H(d_i) \neq H(d_j)) \approx e^{-\sum_{i=0}^{s-1}\frac{i}{2^m}}.$$

The sum of an arithmetic progression

$$S' = \sum_{i=1}^{s-1}\frac{i}{2^m}$$

equals $\frac{s(s-1)}{2}$, thus:

$$P(H(d_i) \neq H(d_j)) \approx e^{-\frac{s(s-1)}{2 \cdot 2^m}},$$

from which we directly obtain

$$P(H(d_i) = H(d_j)) \approx 1 - e^{-\frac{s^2}{2 \cdot 2^m}}. \tag{4}$$

In practice, the so-called "birthday bound" is used. To derive this, let's find the number of samples $k$ needed to achieve a given probability $p$ of collision in a system with $2^m$ possible outcomes.

Using expression (4), solving the equation with respect to $k$, we get:

$$p \approx 1 - e^{-\frac{k(k-1)}{2 \cdot 2^m}},$$

$$e^{-\frac{k(k-1)}{2 \cdot 2^m}} \approx 1 - p,$$

$$-\frac{k(k-1)}{2 \cdot 2^m} \approx \ln(1-p),$$

$$k(k-1) \approx -2 \cdot 2^m \ln(1-p).$$

Approximating $k(k-1)$ as $k^2$ for large $k$, we obtain:

$$k^2 \approx -2 \cdot 2^m \ln(1-p),$$

$$k \approx \sqrt{-2 \cdot 2^m \ln(1-p)}.$$

For $p = 0.5$, the equation simplifies to:

$$k \approx \sqrt{-2 \cdot 2^m \ln(0.5)} = \sqrt{2 \cdot 2^m \ln(2)}.$$

Since $\ln(2)$ is approximately 0.693, the equation becomes:

$$k \approx 1.1774\sqrt{2^m} \approx 2^{m/2}.$$

Thus, a hash length twice as large is required to ensure computational complexity comparable to that of a full brute-force search. In other words, if an attacker can compute $2^{m/2}$ hash values through brute force, they will start finding hash collisions for all hashes shorter than $m$ bits.

In the context of our task to estimate collision probabilities in Merkle Trees when implementing a "birthday attack," in formulas (1) and (2), the value $2^{-m}$ should be replaced with $2^{-m/2}$, i.e., we have:

$$\begin{aligned} P(r=r') &= 2^{-m/2} + (1-2^{-m/2})(1-(1-2^{-m/2})^k) \approx \\ &\approx 2^{-m/2} + e^{-2^{-m/2}}(1-e^{-k2^{-m/2}}) = 2^{-m/2} + e^{-2^{-m/2}} - e^{-(k+1)2^{-m/2}}. \end{aligned} \quad (5)$$

Therefore, the "birthday paradox" strategy leads to a halving of the "effective" hash code length. In the context of our problem, this means that a significantly shorter path length $k$ will be required for a critical increase in collision probability. For example, for the same parameters as in Figure 1, the probability $P(r=r')$ will increase quadratically faster. Specifically, for $m=4$, the probability $P(r=r')$ will reach 0.63 at $k = 2^{m/2} = 4$. This change, compared to the baseline scenario from Figure 1 (lower surface), is presented in Figure 3. For clarity, the dependencies in Figure 3 are shown in different projections.

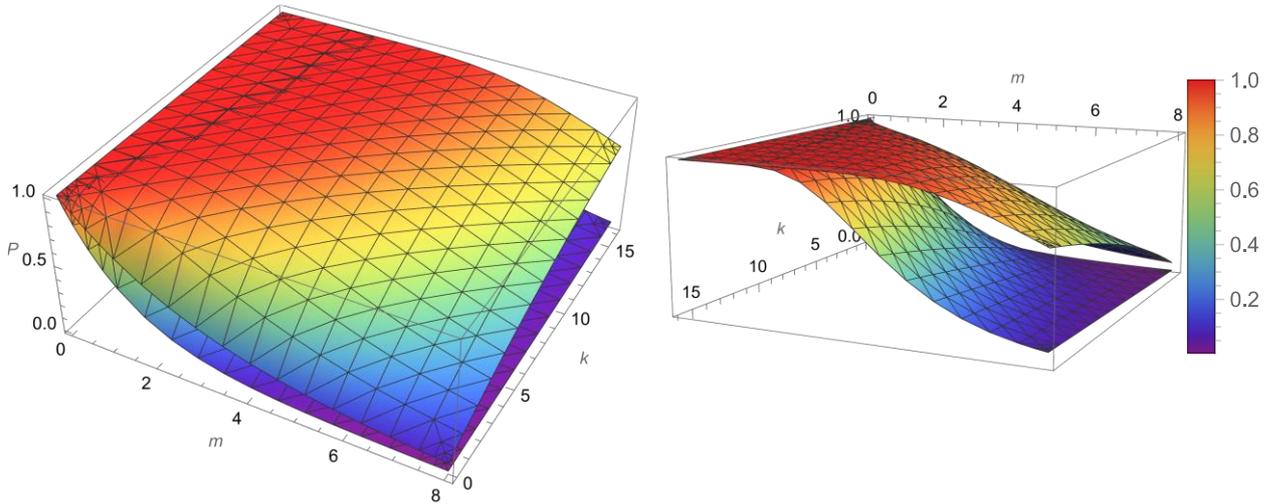

Figure 3 – Dependency of the probability $P(r=r')$ on parameters $k$ and $m$ for the "birthday paradox" strategy (upper surface) compared to the baseline scenario (lower surface).

As a result of our research, precise formulas and corresponding approximations have been derived to estimate the probability of collisions in Merkle Trees. These formulas allow for the consideration of the tree size and the length of the Merkle path, as well as various scenarios of selecting false data to find root collisions. It has been established that increasing the length of the Merkle path leads to a higher probability of collisions, inevitably impacting the security of Merkle Proofs. The most dangerous scenario is an attack based on the "birthday paradox," where any pairs of data leading to root collisions are selected without restrictions on the values of these roots.

At the same time, it should be noted that in practically significant cases, a critical increase in the probability $P(r=r')$ will not occur. For most modern applications, the size of the Merkle Tree is limited to $k \leq 64$. Cryptographically secure hash functions with $m \geq 128$ are used, meaning the expected size of $k$ is significantly lower than $2^m$. For example, in one of the most popular blockchains, Ethereum, the Keccak hashing function with $m=256$ bits is used, and the value of $k$ in Merkle Trees rarely exceeds 56. Referring to formulas (1) and (2), it can be asserted that for these parameters, the probability of collision $P(r=r')$ will remain critically low, posing no substantial threat to security.

### 4.2. Experimental Studies

In the experimental part of our research, we focus on the empirical verification of the theoretical conclusions regarding the probability of root collisions in Merkle Trees. For this purpose, we developed a Python program [24], which allows us to empirically estimate this probability and compare it with theoretical calculations.

**Description of the Experimental Methodology**

1. Hash Generation:
   - The function 'generate_hash' takes input data and hash length, generating a truncated hash using the SHA-256 algorithm [25].
2. Random Data Generation:
   - The function 'generate_random_data' creates random strings of a specified length, used as input data for Merkle Trees.
3. Calculation of Merkle Tree Root Hash:
   - The function 'calculate_merkle_root' computes the root hash for given data and Merkle path, sequentially applying the hashing function.
4. Theoretical Probability:
   - The function 'theoretical_probability' calculates the theoretical probability of root collisions in Merkle Trees based on hash length and path length.
5. Conducting Experiments:
   - The function 'run_experiment' conducts a series of experiments for given values of hash length and Merkle path length. In each experiment, a new Merkle path and root hash are generated, then checked for root hash matches with other randomly generated data.
6. Analysis of Results:
   - The experimental results and theoretical calculations are presented in graph form for a visual comparison. A logarithmic scale is used on the probability axis for better visualization of results at low probabilities.
7. Experimental Parameters:
   - Merkle Path Lengths $k$: Values from 1 to 16 are considered.
   - Hash Lengths $m$: Hashes of 4, 8, 12, and 16 bits (equivalent to 1, 2, 3, and 4 hexadecimal characters) are investigated.
8. Methodology of Conducting Experiments:
   - Using the 'run_experiment' function, a series of experiments are conducted for each combination of $k$ and $m$ values. In each experiment, random data and Merkle paths are generated, followed by the calculation of the root hash.
   - For each set of parameters $k$ and $m$, 1000 experiments (value $N$) are repeated 100 times (value experiments
   - Empirical probabilities of Merkle Tree root collisions are collected and compared with theoretical probabilities calculated based on the proposed formula.

9. Visualization of Results:
   - The experimental results are visualized using graphs, where the X-axis represents the values of $k$, and the Y-axis shows the probabilities of root collisions $P(r=r')$.
   - Separate curves are plotted for each $m$ value, showing the dependency of empirical probability on the length of the Merkle path.
   - A logarithmic scale is used on the Y-axis for better visualization of results, especially at low probabilities.

The experiments are conducted to verify theoretical assumptions about the probability of root collisions in Merkle Trees and to assess the impact of various parameters on this probability. This experimental approach allows us not only to validate theoretical assumptions about the probability of root collisions in Merkle Trees but also to evaluate their practical applicability and reliability under various conditions.

**Experimental Results**

In the experimental part of our study, we conducted a series of experiments to assess the probability of root collisions in Merkle Trees. These experiments aimed to verify the theoretical calculations presented in previous sections and to evaluate the impact of various parameters, such as hash length and Merkle path length, on this probability.

The results of the experiments are vividly presented in Figure 4.

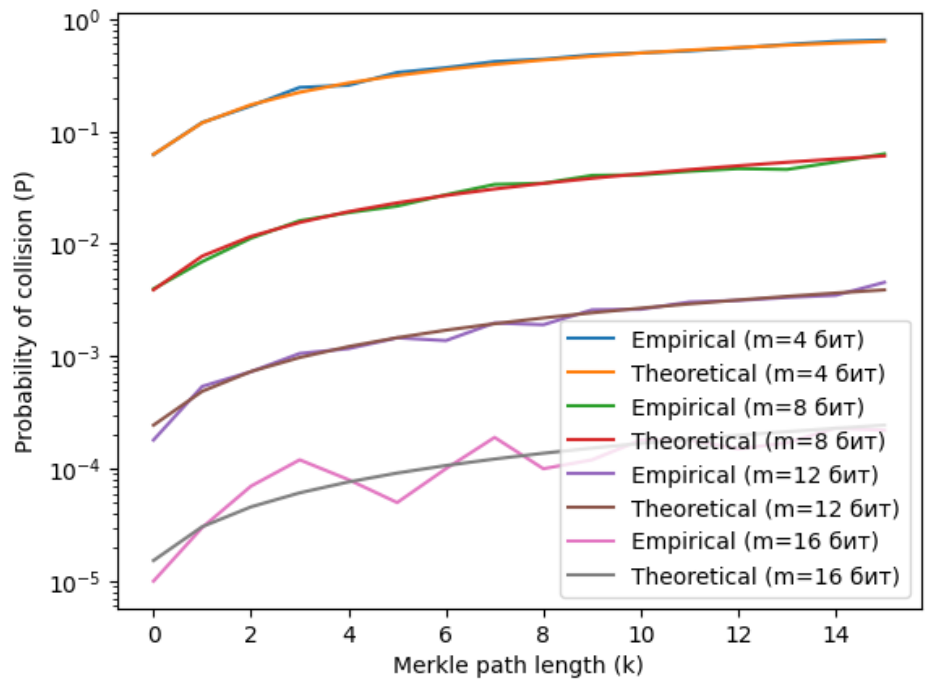

Figure 4 – Experimental Results

Analyzing the data leads to the following conclusions:
1. Convergence of Theoretical Calculations and Experimental Data:
   - The observed convergence between theoretical calculations and experimental results validates our theoretical assumptions. This demonstrates that the proposed mathematical model adequately describes the behavior of Merkle Trees in the context of root collision probability. Notable variations in empirical data for $m=16$ bits are due to the insufficient sample size, as the estimated probabilities are already quite small.
2. Impact of Hash Length and Path Length:

- The results show that with an increase in hash length, the probability of root collisions exponentially decreases, aligning with expectations and emphasizing the importance of choosing a sufficient hash length for security.
- Conversely, as the path length in the Merkle Tree increases, the probability of root collisions also increases, indicating potential security risks when using long paths in Merkle Trees.

3. Visualization of Results:
   - Graphs comparing theoretical and experimental probabilities effectively illustrate the similarity of these data. The logarithmic scale on the probability axis allowed for efficient visualization of results, even at low probability values.

Thus, the experimental results confirm our theoretical calculations and provide important practical insights for the design and use of Merkle Trees in various applications. Particularly significant is the finding that increasing the path length in the Merkle Tree raises the probability of root collisions, potentially reducing system security. These results underscore the necessity of careful parameter selection in Merkle Trees to ensure an optimal balance between efficiency and security in blockchain technologies and other systems where data verification is used.

## 5. Discussion of Research Findings

Our research has yielded significant results that contribute to the understanding of the dynamics and security of Merkle Trees. Theoretical calculations, corroborated by experimental data, have revealed important aspects influencing the probability of root collisions in Merkle Trees, directly relating to their security and efficiency in blockchain systems and other applications.

The theoretical analysis demonstrated that the probability of root collisions in Merkle Trees depends on the hash length and the path length in the tree. Particularly important is the finding that increasing the hash length decreases the probability of collision, underscoring the significance of choosing a sufficient hash length for security. On the other hand, increasing the path length in the Merkle Tree leads to a higher probability of root collisions, which could pose potential security risks for the system.

The experimental results confirmed these theoretical assumptions, showing a convergence between theoretical calculations and practical observations. This emphasizes the reliability of our mathematical approach and its suitability for analyzing Merkle Trees. The visualization of results on graphs vividly demonstrates the similarity between theoretical and experimental data, reinforcing confidence in our conclusions.

Especially crucial is the understanding that finding a balance between efficiency and security is a key aspect in designing and using Merkle Trees. Our results provide valuable recommendations for the practical application of these structures, particularly regarding the choice of optimal hash length and limiting the path length in the tree to ensure security without significantly reducing system performance.

In conclusion, our research opens new perspectives for further studies in optimizing data structures used in blockchain technologies and other systems requiring reliable data verification. The knowledge gained can be used to enhance the security and efficiency of systems employing Merkle Trees, making a significant contribution to the development of modern data processing technologies.

## 6. Conclusion

In our study, we conducted a theoretical analysis and experimental verification of the probability of root collisions in Merkle Trees. The results confirmed that the probability of root collisions depends on the hash length and the path length in the tree, having important practical implications for the security of blockchain systems and other applications using Merkle Trees. We found that increasing the hash length decreases the probability of collision, while increasing the path length in the tree increases this probability. These findings highlight the need for careful selection of Merkle Tree parameters to ensure an optimal balance between efficiency and security. Our research significantly contributes to understanding the dynamics and security of Merkle Trees, providing valuable recommendations for their practical application and further research.

## 7. Statements and Declarations

**Author Contributions:**
- Conceptualization and methodology, Writing-review and editing, Oleksandr Kuznetsov;
- Supervision, Resources; Alex Rusnak;
- Investigation, Anton Yezhov;
- Software and validation, Kateryna Kuznetsova;
- Visualization; Formal analysis, Dzianis Kanonik;
- Validation, Methodology; Oleksandr Domin.

All authors have read and agreed to the published version of the manuscript.

**Data availability**
- The datasets generated during and/or analyzed during the current study are available from the corresponding author on reasonable request.

**Declaration of interests**
- I declare that the authors have no competing financial interests, or other interests that might be perceived to influence the results and/or discussion reported in this paper.
- The results/data/figures in this manuscript have not been published elsewhere, nor are they under consideration (from you or one of your Contributing Authors) by another publisher.
- All of the material is owned by the authors and/or no permissions are required.

**Compliance with ethical standards**
- Mentioned authors have no conflict of interest in this article. This article does not contain any studies with human participants or animals performed by any of the authors.

**Funding:**
This project has received funding from the Proxima Labs, 1501 Larkin Street, suite 300, San Francisco, USA## 8. References

path length in the tree increases this probability. These findings highlight the need for careful selection of Merkle Tree parameters to ensure an optimal balance between efficiency and security. Our research significantly contributes to understanding the dynamics and security of Merkle Trees, providing valuable recommendations for their practical application and further research.

## 7. Statements and Declarations

**Author Contributions:**
- Conceptualization and methodology, Writing-review and editing, Oleksandr Kuznetsov;
- Supervision, Resources; Alex Rusnak;
- Investigation, Anton Yezhov;
- Software and validation, Kateryna Kuznetsova;
- Visualization; Formal analysis, Dzianis Kanonik;
- Validation, Methodology; Oleksandr Domin.

All authors have read and agreed to the published version of the manuscript.

**Data availability**
- The datasets generated during and/or analyzed during the current study are available from the corresponding author on reasonable request.

**Declaration of interests**
- I declare that the authors have no competing financial interests, or other interests that might be perceived to influence the results and/or discussion reported in this paper.
- The results/data/figures in this manuscript have not been published elsewhere, nor are they under consideration (from you or one of your Contributing Authors) by another publisher.
- All of the material is owned by the authors and/or no permissions are required.

**Compliance with ethical standards**
- Mentioned authors have no conflict of interest in this article. This article does not contain any studies with human participants or animals performed by any of the authors.

**Funding:**

This project has received funding from the Proxima Labs, 1501 Larkin Street, suite 300, San Francisco, USA

## 8. References

[1] Z. Xu, L. Wei, J. Wu, C. Long, A Blockchain-Based Digital Copyright Protection System with Security and Efficiency, in: K. Xu, J. Zhu, X. Song, Z. Lu (Eds.), Blockchain Technology and Application, Springer, Singapore, 2021: pp. 34–49. https://doi.org/10.1007/978-981-33-6478-3_3.

[2] Y. Zhang, J. Wang, X. He, J. Liu, Blockchain-Based Access Control Mechanism in Electronic Evidence, in: K. Xu, J. Zhu, X. Song, Z. Lu (Eds.), Blockchain Technology and Application, Springer, Singapore, 2021: pp. 17–33. https://doi.org/10.1007/978-981-33-6478-3_2.

[3] H. Arslanian, Ethereum, in: H. Arslanian (Ed.), The Book of Crypto: The Complete Guide to Understanding Bitcoin, Cryptocurrencies and Digital Assets, Springer International Publishing, Cham, 2022: pp. 91–98. https://doi.org/10.1007/978-3-030-97951-5_3.

[4] J. Rosa-Bilbao, J. Boubeta-Puig, Chapter 15 - Ethereum blockchain platform, in: R. Pandey, S. Goundar, S. Fatima (Eds.), Distributed Computing to Blockchain, Academic Press, 2023: pp. 267–282. https://doi.org/10.1016/B978-0-323-96146-2.00006-1.